\documentclass[twocolumn,superscriptaddress]{quantumarticle}
\usepackage[utf8]{inputenc}
\usepackage{amsmath}
\usepackage{amssymb}
\usepackage[normalem]{ulem}
\usepackage{graphicx}
\usepackage{pdfpages}
\usepackage{color}
\usepackage{float} 
\usepackage{hyperref}
\usepackage{cleveref}

\usepackage[numbers,sort&compress]{natbib}

\usepackage{pdfpages}
\makeatletter
\AtBeginDocument{\let\LS@rot\@undefined}
\makeatother

\crefname{equation}{Eq.}{Eqs.}
\Crefname{equation}{Equation}{Equations}
\crefname{figure}{Fig.}{Figs.}
\Crefname{figure}{Figure}{Figures}
\crefname{section}{Sect.}{Sects.}
\Crefname{section}{Section}{Sections}

\newcommand{\ket}[1]{| #1 \rangle}

\newcommand{\expo}[1]{\text{e}^{ #1 }}

\newcommand{\sx}{\hat{\sigma}_x}
\newcommand{\sz}{\hat{\sigma}_z}
\newcommand{\ha}{\hat{a}}
\newcommand{\had}{\hat{a}^\dagger}
\newcommand{\hb}{\hat{b}}
\newcommand{\hU}{\hat{U}}
\newcommand{\hH}{\hat{H}}
\newcommand{\wc}{\omega_r}
\newcommand{\wm}{\omega_m}

\newcommand{\wai}{\omega_{ai}}
\newcommand{\wao}{\omega_{a1}}
\newcommand{\wat}{\omega_{a2}}
\newcommand{\szo}{\hat{\sigma}_{z1}}
\newcommand{\szt}{\hat{\sigma}_{z2}}
\newcommand{\szi}{\hat{\sigma}_{zi}}

\begin{document}

\title{Fast and High-Fidelity Entangling Gate through Parametrically Modulated Longitudinal Coupling}
\author{Baptiste Royer}
\affiliation{Institut quantique and D\'epartment de Physique, Universit\'e de Sherbrooke, 2500 boulevard de l'Universit\'e, Sherbrooke, Qu\'ebec J1K 2R1, Canada}
\author{Arne L. Grimsmo}
\affiliation{Institut quantique and D\'epartment de Physique, Universit\'e de Sherbrooke, 2500 boulevard de l'Universit\'e, Sherbrooke, Qu\'ebec J1K 2R1, Canada}
\author{Nicolas Didier}
\altaffiliation{Current address: Rigetti Quantum Computing, 775~Heinz Avenue, Berkeley, California 94710, USA.}
\affiliation{QUANTIC team, Inria Paris, 2 rue Simone Iff, 75012 Paris, France}
\author{Alexandre Blais}
\affiliation{Institut quantique and D\'epartment de Physique, Universit\'e de Sherbrooke, 2500 boulevard de l'Universit\'e, Sherbrooke, Qu\'ebec J1K 2R1, Canada}
\affiliation{Canadian Institute for Advanced Research, Toronto, Canada}


\begin{abstract}
We investigate an approach to universal quantum computation based on the modulation of longitudinal qubit-oscillator coupling. 
We show how to realize a controlled-phase gate by simultaneously modulating the longitudinal coupling of two qubits to a common oscillator mode. In contrast to the more familiar transversal qubit-oscillator coupling, the magnitude of the effective qubit-qubit interaction does not rely on a small perturbative parameter. As a result, this effective interaction strength can be made large, leading to short gate times and high gate fidelities. We moreover show how the gate infidelity can be exponentially suppressed with squeezing and how the entangling gate can be generalized to qubits coupled to separate oscillators. Our proposal can be realized in multiple physical platforms for quantum computing, including superconducting and spin qubits.
\end{abstract}

\maketitle

\textit{Introduction}\textemdash 
A widespread strategy for quantum information processing is to couple the dipole moment of multiple qubits to common  oscillator modes, the latter being used to measure the qubits and to mediate long-range interactions. Realizations of this idea are found in Rydberg atoms~\cite{haroche:2006a}, superconducting qubits~\cite{Blais:04a} and quantum dots~\cite{imamoglu:1999a} amongst others. With the dipole moment operator being off-diagonal in the qubit's eigenbasis, this type of \emph{transversal} qubit-oscillator coupling leads to hybridization of the qubit and oscillator degrees of freedom. In turn, this results in qubit Purcell decay~\cite{Houck_2008} and to qubit readout that is not truly quantum non-demolition (QND)~\cite{Boissonneault_2009}. To minimize these problems, the qubit can be operated at a frequency detuning from the oscillator that is large with respect to the transverse coupling strength $g_x$. This interaction then only acts perturbatively, taking a dispersive character~\cite{haroche:2006a}. While it has advantages, this perturbative character also results in slow oscillator-mediated qubit entangling gates~\cite{blais:2007a,chow:2013a,Paik:16a}.

Rather than relying on the standard transversal coupling, $H_x=g_x(\had+\ha)\sx$, an alternative approach is to use a longitudinal interaction, $H_z=g_z(\had+\ha)\sz$~\cite{Kerman:08a,Kerman:13a,Billangeon:15a,Billangeon:15b,Didier:15a,Richer:16a}. Since $H_z$ commutes with the qubit's bare Hamiltonian the qubit is not dressed by the oscillator. Purcell decay is therefore absent~\cite{Kerman:13a,Billangeon:15a} and qubit readout is truly QND~\cite{Didier:15a}. The absence of qubit dressing also allows for scaling up to a lattice of arbitrary size with strictly local interactions~\cite{Billangeon:15a}.

By itself, longitudinal interaction however only leads to a vanishingly small qubit state-dependent displacement of the oscillator field of amplitude $g_z/\wc \ll 1$, with $\wc$ the oscillator frequency. In Ref.~\cite{Didier:15a}, it was shown that modulating $g_z$ at the oscillator frequency $\wc$ activates this interaction leading to a large qubit state-dependent oscillator displacement and to fast QND qubit readout. In this paper, we show how the same approach can be used, together with single qubit rotations, for universal quantum computing by introducing a fast and high-fidelity controlled-phase gate based on longitudinal coupling. The two-qubit logical operation relies on parametric modulation of a longitudinal qubit-oscillator coupling, inducing an effective $\sz\sz$ interaction between qubits coupled to the same mode. A similar gate was first studied in Ref.~\cite{Kerman:13a} in the presence of an additional dispersive interaction $\chi \had \ha \sz$ and a cavity drive. We show that, with a purely longitudinal interaction excluding the former term, the gate fidelity can be improved exponentially using squeezing. We moreover show that the gate can be performed remotely on qubits coupled to separate but interacting oscillators. The latter allows for a modular architecture that relaxes design constraints and avoids spurious interactions while maintaining minimal circuit complexity~\cite{Billangeon:15a,Richer:16a,Brecht:16a}.

In contrast to two-qubit gates based on a transversal interaction~\cite{blais:2007a,chow:2013a}, this proposal does not rely on strong qubit-oscillator detuning and is not based on a perturbative argument. As a result, the longitudinally mediated $\sz\sz$ interaction is valid for all qubit, oscillator and modulation parameters and does not result in unwanted residual terms in the Hamiltonian. For this reason, in the ideal case where the interaction is purely longitudinal (i.e. described by $H_z$), there are no fundamental bounds on gate infidelity or gate time and both can in principle be made arbitrarily small simultaneously.

Similarly to other oscillator-mediated gates, loss from the oscillator during the gate leads to gate infidelity. This can be minimized by working with high-Q oscillators something that is, however, in contradiction with the requirements for fast qubit readout~\cite{Blais:04a}. We solve this dilemma by exploiting quantum bath engineering, using squeezing at the oscillator input. By appropriately choosing the squeezed quadrature, we show how `which-qubit-state' information carried by the photons leaving the oscillator can be erased. This leads to an exponential improvement in gate fidelity with squeezing strength. 

\textit{Oscillator mediated qubit-qubit interaction}\textemdash
Following Ref.~\cite{Didier:15a}, we consider two qubits coupled to a single harmonic mode via their $\sz$ degree of freedom. Allowing for a time-dependent coupling, the Hamiltonian reads ($\hbar = 1$) 
\begin{equation}\label{H}
\begin{aligned}
\hH(t) =\;& \wc \ha^{\dag} \ha + \tfrac{1}{2}\wao\szo + \tfrac{1}{2}\wat\szt\\
&+ g_{1}(t)\szo\; (\ha^{\dag} + \ha) + g_{2}(t) \szt\;(\ha^{\dag} + \ha).
\end{aligned}
\end{equation}
In this expression, $\wc$ and $\wai$ are the frequencies of the oscillator and of the $i^{\text{th}}$ qubit, respectively, while $g_{i}(t)$ are the corresponding longitudinal coupling strengths.

For constant couplings, $g_i(t) = g_i$, the longitudinal interaction only leads to a displacement of order $\sim g_i/\omega_r$, which is vanishingly small for typical parameters.
This interaction can be rendered resonant by modulating $g_i(t)$ at the oscillator frequency leading to a large qubit-state dependent displacement of the oscillator state. Measurement of the oscillator by homodyne detection can then be used for fast QND qubit readout~\cite{Didier:15a}. Consequently, modulating the coupling at the oscillator frequency rapidly dephases the qubits. To keep dephasing to a minimum, we instead use an off-resonant modulation of  $g_{i}(t)$ at a frequency $\wm$ detuned from $\wc$ by many oscillator linewidths $\kappa$: $g_i(t) = g_i \cos(\wm t)$, where $g_{1,2}$ are constant real amplitudes~\cite{Kerman:13a}.

The oscillator-mediated qubit-qubit interaction can be made more apparent by applying a polaron transformation $\hU(t) = \exp[\sum_{i=1,2}\alpha_i(t) \szi \ha^\dag - {\rm H.c.}]$ with an appropriate choice of $\alpha_{i}(t)$ (see supplemental material). Doing this, we find in the polaron frame the simple Hamiltonian
\begin{equation}\label{Hdisplaced}
  \hH_\mathrm{pol}(t) = \wc \ha^\dag \ha + J_{z}(t) \szo\szt.
\end{equation}
The full expression for the $\sz\sz$-coupling strength $J_z(t)$ is given in the supplemental material. In the following we will, however, assume two conditions on the total gate time, $t_g$, such that this expression simplifies greatly. For $\delta t_g = n\times 2\pi$ and $\omega_m t_g = m\times \pi$, with $n$ and $m$ integers, we can replace $J_z(t)$ by
\begin{equation}\label{eq:J_z}
  \bar J_{z} =  -\frac{g_1 g_2}{2} \left[\frac{1}{\delta} {+} \frac{1}{\omega_r+\omega_m} \right],
\end{equation}
where $\delta \equiv \wc -\wm$ is the modulation drive detuning.

By modulating the coupling for a time $t_g = \theta/ 4|\bar J_{z}|$, evolution under Eq.~\eqref{Hdisplaced} followed by single qubit $Z$-rotations leads to the entangling controlled-phase gate $U_{CP}(\theta) = \text{diag}[1,1,1,\expo{i\theta}]$.
Since $U_{CP}(\pi)$ together with single qubit rotations forms a universal set~\cite{Nielsen:00a}, we only consider this gate from now on.

Note that the conditions on the gate time used in~\cref{eq:J_z} are not necessary for the validity of~\cref{Hdisplaced}, and the gate can be realized without these assumptions. However, as we will discuss below, these conditions are important for optimal gate performance: They ensure that the oscillator starts and ends in the vacuum state, which implies that the gate does not need to be performed adiabatically. Finally, not imposing the second constraint, $\omega_m t_g = m\times\pi$, only introduces fast rotating terms to~\cref{eq:J_z} which we find to have negligible effect for the parameters used later in this paper. In other words, this constraint can be ignored under a rotating-wave approximation.

The above situation superficially looks similar to controlled-phase gates based on transversal coupling and strong oscillator driving~\cite{blais:2007a,Cross:15a,Puri:16b}. There are, however, several key differences. With transversal coupling, the $\sz\sz$ interaction is derived using perturbation theory and is thus only approximately valid for small $g_x/\{\Delta,\delta_\mathrm{d}\}$, with $\Delta$ the qubit-oscillator detuning and $\delta_\mathrm{d}$ the oscillator-drive detuning. For the same reason, it is also only valid for small photon numbers $n \ll n_{\rm crit} = \Delta^2 / 4 g_x^2$~\cite{blais:2007a}. Moreover, this interaction is the result of a fourth order process in $g_x/\{\Delta,\delta_\mathrm{d}\}$, leading to slow gates. Because of the breakdown of the dispersive approximation, attempts to speed up the gate by decreasing the detunings or increasing the drive amplitude have resulted in low gate fidelities~\cite{chow:2013a}. In contrast, with longitudinal coupling, the $\sz\sz$ interaction conveniently scales as $\sim g_1g_2/\delta$, \emph{i.e.} it scales as a second-order process in $g_{1,2}/\delta$, but the exact nature of the transformation means that there are no higher order terms. Consequently, \cref{Hdisplaced} is valid for any value of $g_{1,2}/\delta$, independent of the oscillator photon number. As will become clear later, this implies that the gate time and the gate infidelity can be decreased simultaneously. Finally, with longitudinal coupling, there is no constraint on the qubit frequencies, in contrast with usual oscillator-induced phase gates where the detuning between qubits should preferably be small. 

\begin{figure}[!t]
\centering
\includegraphics{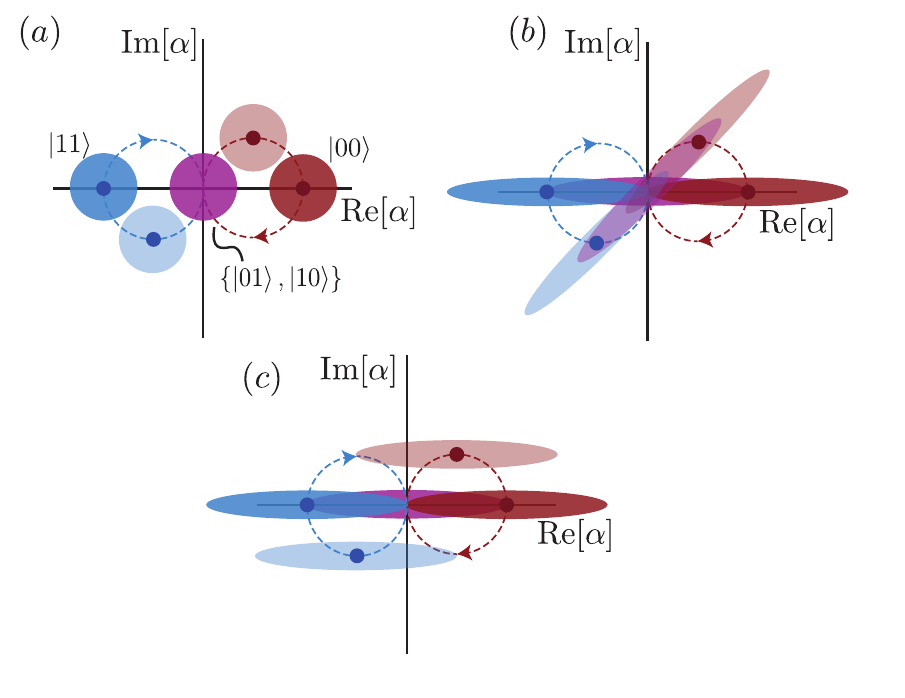}
\caption{Schematic illustration, in a frame rotating at $\wc$, of the qubit-state dependent oscillator field in phase space for $g_1=g_2$ starting and ending in the vacuum state (purple). The oscillator's path for $\ket{00}$ ($\ket{11}$) is shown by the dashed red (blue) line. The qubit-state dependent oscillator state is shown in light ($t=t_g/4$) and dark colors ($t = t_g/2$). The oscillator's state associated to $\{\ket{01},\ket{10}\}$ stays in the vacuum state for the duration of the gate (purple). (a) No squeezing.
(b,c) Squeezing can help in erasing the which-qubit-state information. 
}
\label{flowers}
\end{figure}

\textit{Oscillator-induced qubit dephasing}\textemdash
\cref{flowers} illustrates, for $g_1=g_2$, the mechanism responsible for the qubit-qubit interaction. Under longitudinal coupling, the oscillator field is displaced in a qubit-state dependent way, following the dashed lines in \cref{flowers}(a) (Panels (b) and (c) will be discussed later). This conditional displacement leads to a non-trivial qubit phase accumulation.
This schematic illustration also emphasizes the main cause of gate infidelity for this type of controlled-phase gate, irrespective of its longitudinal or transversal nature: Photons leaking out from the oscillator during the gate carry information about the qubit state, leading to dephasing.

A quantitative understanding of the gate infidelity under photon loss can be obtained by deriving a master equation for the joint qubit-oscillator system. While general expressions are given in the supplemental material, to simplify the discussion we assume here that $g_1 = g_2 \equiv g$. Following the standard approach~\cite{Gardiner:04a}, the Lindblad master equation in the polaron frame reads
\begin{equation}
\label{ME}
\begin{aligned}
  \dot{\rho}(t) ={}& -i[\hH_\mathrm{pol},\rho(t)] + \kappa \mathcal{D}[\ha]\rho(t)\\
  &+ \Gamma[1-\cos(\delta t)] \mathcal{D}[\szo + \szt]\rho(t),
\end{aligned}
\end{equation}
where $\kappa$ is photon decay rate and $\mathcal{D}[x]$ denotes the usual dissipation super-operator $\mathcal{D}[x]\bullet = x\bullet x^\dag - \frac{1}{2}\{x^\dag x,\bullet\}$. 
The last term of \cref{ME} corresponds to a dephasing channel with rate $\Gamma = 2\kappa (g/2\delta)^2 $. 
Since $\hH_\mathrm{pol}$ does not generate qubit-oscillator entanglement during the evolution, we can ensure that in this frame, gate-induced dephasing only happens due to the last term in~\cref{ME} with rate $\Gamma$, by imposing that the initial and final polaron transformations also do not lead to qubit-oscillator entanglement. This translates to the condition $\alpha_i(0) = \alpha_i(t_g) = 0$ and is realized for $\delta t_g = n\times 2\pi$, which is the constraint mentioned earlier (neglecting fast-rotating terms related to the second constraint $\omega_m t_g = m\times\pi$). More intuitively, it amounts to completing $n$ full circles in \cref{flowers}, the oscillator ending back in its  initial unentangled state. Note that these conclusions remain unchanged if the oscillator is initially in a coherent state. As a result, there is no need for the oscillator to be empty at the start of the gate~\cite{Kerman:13a}.

\begin{figure}[!t]
\centering
\includegraphics{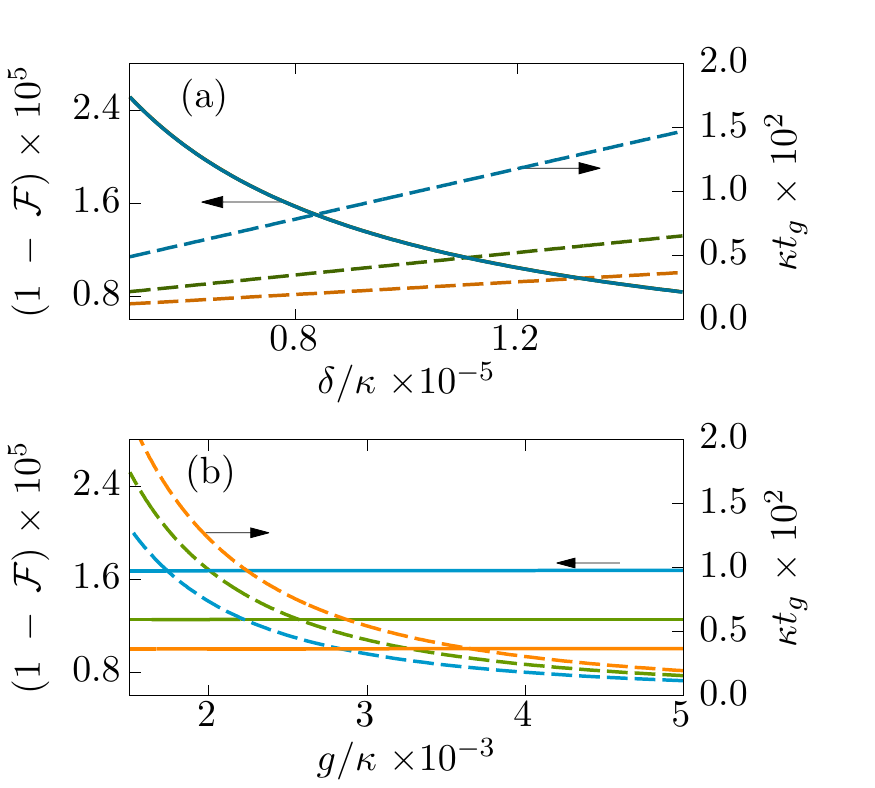}
\caption{Average gate infidelity $1-\mathcal{F}$ (full line) and gate time (dashed lines) of $U_{CP}(\pi)$ as a function of (a) detuning and (b) coupling strength. 
In panel (a) $g/\kappa \times 10^{-3}$ is fixed at 2 (blue), 3 (green), 4 (orange). Note that the corresponding three infidelity curves are indistinguishable on this scale. 
In panel (b) $\delta/\kappa \times 10^{-5}$ is fixed at 0.75 (blue), 1 (green), 1.25 (orange).}
\label{Fid_g}
\end{figure}

Based on the dephasing rate $\Gamma$ and on the gate time $t_g$, a simple estimate for the scaling 
of the gate infidelity is $1-\mathcal{F} \sim \Gamma \times t_g \sim \kappa/\delta$ ~\footnote{Note that $1-\mathcal{F}$ refers only to the error due to photon decay, excluding the qubits natural $T_1$ and $T_2$ times.}. A key observation is that this gate error is independent of $g$, while the gate time scales as $t_g \sim \delta/g^2$. 
Both the gate time and the error can therefore, in principle, be made arbitrarily small simultaneously. This scaling of the gate error and gate time is confirmed by the numerical simulations of \cref{Fid_g}, which shows the dependence of the gate infidelity~\cite{Nielsen:02a} on detuning $\delta$ and coupling strength $g$, as obtained from numerical integration of \cref{ME}.
 The expected increase in both fidelity (full lines) and gate time (dashed lines) with increasing detuning $\delta$ are apparent in panel (a). In addition, panel (b) confirms that, to a very good approximation, the fidelity is independent of $g$ (full lines) while the gate time decreases as $t_g \sim 1/g^2$ (dashed lines). 

This oscillator-induced phase gate can be realized in a wide range of physical platforms where longitudinal coupling is possible. Examples include spin qubits in inhomogeneous magnetic field~\cite{Beaudoin:16a}, singlet-triplet spin qubits~\cite{Jin:12a}, flux qubits capacitively coupled to a resonator~\cite{Billangeon:15a} and transmon-based superconducting qubits \cite{Didier:15a,Richer:16a,Kerman:13a}.
The parameters used in \cref{Fid_g} have been chosen following the latter references.
In particular, taking $\kappa/2\pi = 0.05$ MHz~\cite{Bruno:15a}, $g/2\pi = 60$ MHz~\cite{Didier:15a} and $\delta/2\pi = 537$ MHz results in a very short gate time of $t_g =$ 37 ns with an average gate infidelity as small as 1 $\times 10^{-4}$.
Taking into account finite qubit lifetimes $T_1=30\,\mu$s and $T_2=20\,\mu$s~\cite{Corcoles:15a}, we find that the infidelity is increased to $\sim 10^{-3}$ (see supplemental material). In other words, the gate fidelity is limited by the qubit's natural decoherence channels with these parameters. For a comparison with transversal resonator-induced phase gate, see the supplemental material.

A crucial feature of this gate is that the circular path followed by the oscillator field in phase space maximizes qubit-state dependent phase accumulation while minimizing dephasing, allowing for high gate fidelities. In contrast to~\cite{Kerman:13a}, this relies on the assumption that there is no dispersive interaction of the form $\chi \had \ha \sz$ in~\cref{H}. Furthermore, we show below that this also allows for exponential improvement in gate fidelity with squeezing. It is therefore desirable to minimize, or avoid completely, dispersive coupling in experimental implementations~\footnote{In the proposal of Ref.~\cite{Didier:15a}, this can be done by reducing the participation ratio, $\eta$, such that $\chi \lesssim \kappa$}.

\textit{Improved fidelity with squeezing}\textemdash
\begin{figure}[!t]
\centering
\includegraphics{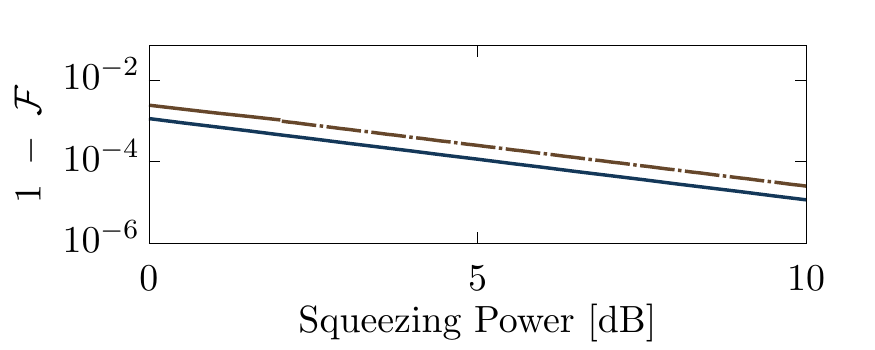}
\caption{Average gate infidelity $1-\mathcal{F}$ squeezing power. Parameters are $\delta/2\pi$ = 0.6 GHz, $g/2\pi$ = 60 MHz, $t_g$ = 42.7 ns, $\kappa/2\pi$ = 1 MHz. In brown, rotating squeezing angle as illustrated in~\cref{flowers}(b). In dark blue, squeezing at $\wc$ as illustrated in~\cref{flowers}(c), and $\kappa(\wm) = 0$ simulating a filter reducing the density of modes to zero at $\wm$.}
\label{Fid_squeezing}
\end{figure}
As discussed above, for fixed $g$ and $\delta$ the fidelity increases with decreasing $\kappa$. A small oscillator decay rate $\kappa$, however, comes at the price of longer measurement time if the same oscillator is to be used for readout~\cite{Didier:15a}.
This problem can be solved by sending squeezed radiation to the oscillator's readout port.
As schematically illustrated in \cref{flowers}, by orienting the squeezing axis with the direction of the qubit-dependent displacement of the oscillator state, the which-path information carried by the photons leaving the oscillator can be erased. By carefully choosing the squeezing angle and frequency, it is thus possible to improve the gate performance without reducing $\kappa$. We now show two different approaches to realize this, referring the reader to the supplemental material for technical details.

A first approach is to send broadband two-mode squeezed vacuum at the input of the oscillator, where the squeezing source is defined by a pump frequency $\omega_p = (\wc+\wm)/2$ and a squeezing spectrum with large degree of squeezing at $\wc$ and $\wm$. A promising source of this type of squeezing is the recently developed Josephson travelling wave amplifiers~\cite{Macklin:15a,White15}. With such a squeezed input field, a coherent state of the oscillator becomes a squeezed state with a squeezing angle that rotates at a frequency $\delta/2$. As illustrated in \cref{flowers}(b), this is precisely the situation where the anti-squeezed quadrature and the displacement of the oscillator's state are aligned at all times. This leads to an exponential decrease in dephasing rate
\begin{equation}\label{gamma}
\Gamma(r)\sim\expo{-2r}\Gamma(0),
\end{equation}
with $r$ the squeezing parameter. This reduction in dephasing rate leads to the exponential improvement in gate fidelity with squeezing power shown by the brown line in \cref{Fid_squeezing}(c). An interesting feature in this Figure is that increasing $\kappa$ by 2 orders of magnitude to allow for fast measurement~\cite{Didier:15a}, leads to the same $\sim 10^{-5}$ gate infidelity obtained above without squeezing here using only $\sim$6 dB of squeezing.
Since numerical simulations are intractable for large amount of squeezing, we depict the infidelity obtained from a master equation simulation by a solid line and the expected infidelity from analytical calculations by a dash-dotted line.

An alternative solution is to use broadband squeezing centered at the oscillator's frequency, \emph{i.e.}~a squeezing source defined by a pump frequency $\omega_p=\wc$. As illustrated in \cref{flowers}(c), using this type of input leads to a squeezing angle that is constant in time in a frame rotating at $\wc$. With this choice, information about the qubits' state contained in the $\ha^\dagger+\ha$ quadrature of the field is erased while information in the $i(\ha^\dagger-\ha)$ quadrature is amplified (cf.~\cref{flowers}). By itself, this does not lead to a substantial fidelity improvement. However, a careful treatment of the master equation shows that \cref{gamma} can be recovered by adding a filter reducing the density of modes at $\wm$ to zero at the output port of the oscillator (see supplemental material). 
Filters of this type are routinely used experimentally to reduce Purcell decay of superconducting qubits~\cite{Reed_2010,Bronn:15a}. As illustrated by the dark blue line in \cref{Fid_squeezing}(c), using single-mode squeezing at $\wc$ and a filter at the modulation frequency, we recover the same exponential improvement found with two-mode squeezing, \cref{gamma}, in addition to a factor of two decrease in gate infidelity without squeezing.

Interestingly, rotating the squeezing axis by $\pi/2$ when squeezing at the oscillator frequency helps in distinguishing the different oscillator states and has been shown to lead to an exponential increase in the signal-to-noise ratio for qubit readout~\cite{Didier:15a}. In practice, the difference between performing a two-qubit gate and a measurement is thus the parametric modulation frequency (off-resonant for the gate and on resonance for measurement) and the choice of squeezing axis.

We note that \cref{gamma} was derived from a master equation treatment under the standard secular approximation~\cite{Breuer:07a}, which is not valid at high squeezing powers (here, $\gtrsim 10$ dB, see supplemental material). At such high powers, the frequency dependence of $\kappa$ together with other imperfections are likely to be relevant.

\textit{Scalability}\textemdash
So far we have focused on two qubits coupled to a single common oscillator. As shown by Billangeon~\emph{et al.}~\cite{Billangeon:15a}, longitudinal coupling of several qubits to separate oscillators that are themselves coupled transversely has favorable scaling properties. Circuits implementing this idea were also proposed by Richer \emph{et al.}~\cite{Richer:16a}.
Interestingly, the gate introduced in this paper can also be implemented in such an architecture.
Consider two qubits interacting with distinct, but coupled, oscillators with the corresponding Hamiltonian~\cite{Billangeon:15a}
\begin{equation}\label{Hscalable}
\begin{aligned}
\hH_{ab} =\;& \omega_a \ha^{\dag} \ha + \omega_b \hb^{\dag} \hb + \tfrac{1}{2}\wao\szo + \tfrac{1}{2}\wat\szt\\
&+ g_{1}(t)\szo\; (\ha^{\dag} + \ha) + g_{2}(t) \szt\;(\hb^{\dag} + \hb)\\
&- g_{ab}(\ha^{\dag} - \ha)(\hb^{\dag} - \hb).
\end{aligned}
\end{equation}
In this expression, $\ha$, $\hb$ label the mode of each oscillator of respective frequencies $\omega_{a,b}$, and $g_{ab}$ is the oscillator-oscillator coupling. As above, $g_{1,2}(t)$ are modulated at the same frequency $\wm$, corresponding to the detunings $\delta_a \equiv \omega_a - \wm$ and $\delta_b \equiv \omega_b - \wm$.
Following the same procedure as above and performing a rotating-wave approximation for simplicity, we find a Hamiltonian in the polaron frame of the same form as \cref{Hdisplaced}, but now with a modified $\sz\sz$ interaction strength
\begin{equation}\label{ZZscalable}
  \bar J_{z} = \frac{1}{2} \frac{g_1 g_2 g_{ab}}{\bar{\delta}^2 - g_{ab}^2(1+\zeta^2)},
\end{equation}
where $\bar{\delta} = (\delta_a+\delta_b)/2$ and $\zeta = (\omega_b - \omega_a)/(2g_{ab})$. 
This implementation allows for a modular architecture, where each unit cell is composed of a qubit and coupling oscillators, used for both readout and entangling gates. 
Such a modular approach can relax design constraints and avoids spurious interactions with minimal circuit complexity~\cite{Billangeon:15a,Richer:16a,Brecht:16a}.

\textit{Conclusion}\textemdash
We have proposed a controlled-phase gate based on purely longitudinal coupling of two qubits to a common oscillator mode. The key to activating the qubit-qubit interaction is a parametric modulation of the qubit-oscillator coupling at a frequency far detuned from the oscillator. The gate infidelity and gate time can in principle be made arbitrarily small simultaneously, in stark contrast to the situation with transversal coupling. We have also shown how the gate fidelity can be exponentially increased using squeezing and that it is independent of qubit frequencies. The gate can moreover be performed remotely in a modular architecture based on qubits coupled to separate oscillators.
Together with the fast, QND and high-fidelity measurement scheme presented in Ref.~\cite{Didier:15a}, this makes a platform based on parametric modulation of longitudinal coupling a promising path towards universal quantum computing in a wide variety of physical realizations.

\begin{acknowledgments}
We thank J.~Bourassa, D.~Poulin and S.~Puri for useful discussions. This work was supported by the Army Research Office under Grant No. W911NF-14-1-0078 and NSERC. This research was undertaken thanks in part to funding from the Canada First Research Excellence Fund and the Vanier Canada Graduate Scholarships. 
\end{acknowledgments}

\bibliographystyle{apsrev4-1}

%

\clearpage


\section*{Supplementary material}

# Supplemental Material for "Fast and High-Fidelity Entangling Gate through Parametrically Modulated Longitudinal Coupling"

Baptiste Royer[1], Arne L. Grimsmo[1], Nicolas Didier[2], and Alexandre Blais[1][3]

[1]Institut quantique and Départment de Physique, Université de Sherbrooke, 2500 boulevard de l'Université, Sherbrooke, Québec J1K 2R1, Canada

[2]Centre de Recherche Inria de Paris, 2 rue Simone Iff, 75012 Paris, France

[3]Canadian Institute for Advanced Research, Toronto, Canada

May 9, 2017

This supplemental material is organized as follows: In Sect. 1 we derive a general master equation for two qubits coupled to a single oscillator that is itself coupled to an external bath. We then use this result to study three cases: a standard bath at zero temperature, a bath that is two-mode squeezed at the modulation and oscillator frequency and a bath that is squeezed at the oscillator frequency. In Sect. 2 we give more information about possible physical implementations. In Sect. 3 we show how the gate can be realized for qubits that are in separate, but coupled, oscillators. In Sect. 4 we give more details about the numerical simulations, followed by a comparison with a transversal resonator-induced phase gate in Sect. 5 and finally in Sect. 6 we derive an error bound for the secular approximation made in deriving the master equation.

## 1  Derivation of the master equation

In this section, we follow the procedure outlined in the main paper, taking damping of the oscillator into account. We start with the full Hamiltonian of two qubits longitudinally coupled to an oscillator, and a bath coupled to the oscillator ($\hbar = 1$),

$$\hat{H} = \hat{H}_0 + \hat{H}_{qr} + \hat{H}_{rf}, \tag{S1}$$

$$\hat{H}_0 = \omega_r \hat{a}^\dagger \hat{a} + \tfrac{1}{2}\omega_{a1}\hat{\sigma}_{z1} + \tfrac{1}{2}\omega_{a2}\hat{\sigma}_{z2} + \int_0^\infty d\omega\, \omega \hat{b}_\omega^\dagger \hat{b}_\omega, \tag{S2}$$

$$\hat{H}_{qr} = g_1(t)\hat{\sigma}_{z1}\,(\hat{a}^\dagger + \hat{a}) + g_2(t)\hat{\sigma}_{z2}\,(\hat{a}^\dagger + \hat{a}), \tag{S3}$$

$$\hat{H}_{rf} = \int_0^\infty \frac{d\omega}{\sqrt{2\pi}}\sqrt{\kappa(\omega)}(\hat{a}+\hat{a}^\dagger)(\hat{b}_\omega + \hat{b}_\omega^\dagger), \tag{S4}$$

where $\hat{H}_{qr}$ is the qubit-oscillator coupling Hamiltonian and $\hat{H}_{rf}$ is the oscillator-bath coupling Hamiltonian. $\hat{b}_\omega$ ($\hat{b}_\omega^\dagger$) is a bath mode annihilation (creation) operator, satisfying the commutation relation $[\hat{b}_\omega, \hat{b}_{\omega'}^\dagger] = \delta(\omega - \omega')$, and $\kappa(\omega)$ is the damping rate of the oscillator at frequency $\omega$. We assume the form $g_1(t) = g_1\cos(\omega_m t)$, $g_2(t) = g_2\cos(\omega_m t + \phi)$ for the qubit-oscillator couplings, with $\phi = 0$ and the modulation frequency far from the oscillator frequency $\delta \equiv \omega_r - \omega_m$. Setting $\phi = \pi$ leads to a very similar derivation, with ultimately a sign difference in the $\hat{\sigma}_{z1}\hat{\sigma}_{z2}$ interaction and allowing us to choose between a ferromagnetic or an antiferromagnetic interaction.

Following the approach outlined in the paper, we first go to a polaron frame by applying the unitary transformation

$$\hat{U}(t) = \exp\left[(\alpha_1\hat{\sigma}_{z1} + \alpha_2\hat{\sigma}_{z2})\,\hat{a}^\dagger - (\alpha_1^*\hat{\sigma}_{z1} + \alpha_2^*\hat{\sigma}_{z2})\,\hat{a}\right], \tag{S5}$$

1

with
$$\alpha_j(t) = \frac{g_j}{2}\left(\frac{e^{-i\omega_m t} - e^{-i\omega_r t}}{\delta} + \frac{e^{i\omega_m t} - e^{-i\omega_r t}}{\omega_r + \omega_m}\right), \tag{S6}$$

leading to the transformed Hamiltonian
$$\begin{aligned}
\hat{H}_{\text{pol}} &= \hat{\mathbf{U}}^\dagger \hat{H} \hat{\mathbf{U}} - i\hat{\mathbf{U}}^\dagger \dot{\hat{\mathbf{U}}} \\
&= \omega_r \hat{a}^\dagger \hat{a} + J_z(t)\hat{\sigma}_{z1}\hat{\sigma}_{z2} + \tfrac{1}{2}\omega_{a1}\hat{\sigma}_{z1} + \tfrac{1}{2}\omega_{a2}\hat{\sigma}_{z2} \\
&\quad + \int_0^\infty \frac{d\omega}{\sqrt{2\pi}}\sqrt{\kappa(\omega)}(\hat{a} + \hat{a}^\dagger)(\hat{b}_\omega e^{-i\omega t} + \hat{b}_\omega^\dagger e^{i\omega t}) \\
&\quad - \int_0^\infty \frac{d\omega}{\sqrt{2\pi}}\sqrt{\kappa(\omega)}(\hat{\mathcal{O}}e^{-i\omega_m t} + \hat{\mathcal{O}}e^{i\omega_m t})(\hat{b}_\omega e^{-i\omega t} + \hat{b}_\omega^\dagger e^{i\omega t}) \\
&\quad + \int_0^\infty \frac{d\omega}{\sqrt{2\pi}}\sqrt{\kappa(\omega)}(\hat{\mathcal{O}}e^{-i\omega_r t} + \hat{\mathcal{O}}e^{i\omega_r t})(\hat{b}_\omega e^{-i\omega t} + \hat{b}_\omega^\dagger e^{i\omega t}).
\end{aligned} \tag{S7}$$

To simplify expressions, we have defined a two-qubit operator
$$\hat{\mathcal{O}} \equiv \frac{g_1\hat{\sigma}_{z1} + g_2\hat{\sigma}_{z2}}{2}\left(\frac{1}{\delta} + \frac{1}{\omega_r + \omega_m}\right) \equiv \frac{g_1\hat{\sigma}_{z1} + g_2\hat{\sigma}_{z2}}{2\tilde{\delta}}, \tag{S8}$$

with $\tilde{\delta} = [1/\delta + 1/(\omega_r + \omega_m)]^{-1}$, and the qubit-qubit coupling strength
$$\begin{aligned}
J_z(t) = &-\frac{g_1 g_2}{2\delta}\left\{1 - \cos(\delta t) - \cos[(\omega_m + \omega_r)t] + \frac{2\omega_r}{\omega_r + \omega_m}\cos(2\omega_m t)\right\} \\
&- \frac{g_1 g_2}{2(\omega_m + \omega_r)}\left\{1 - \cos(\delta t) - \cos[(\omega_m + \omega_r)t]\right\}.
\end{aligned} \tag{S9}$$

It is important to note that, up to this point, all transformations performed are exact. Also, we have specifically chosen $\alpha(t)$ in such a way that at all times the oscillator's state is vacuum in the polaron frame: We will use this fact later in the discussion.

In practice Eq. (S9) can be simplified greatly if we impose the two conditions $\delta t_g = 2n\pi$, $2\omega_m t_g = 2m\pi$ on the modulation parameters, with n and m integers. In this case, all the cosine terms in Eq. (S9) average to zero and can be dropped exactly. We can thus use
$$J_z(t) = \bar{J}_z = -\frac{g_1 g_2}{2\tilde{\delta}} \qquad \text{for } \delta t_g = n \times 2\pi, \quad \omega_m t_g = m \times \pi. \tag{S10}$$

We emphasize that in the main paper we used $\bar{J}_z$ for the qubit-qubit coupling to simplify the discussion, but in general there is no obstacle to keeping the full form of $J_z(t)$, Eq. (S9). In the numerical results presented, we always use the full form.

To derive the Lindblad Master equation, we transform to the interaction picture with the unitary transformation
$$\hat{U}_I = \exp\left\{-i\int_0^t ds\,[H_0 + J_z(s)\hat{\sigma}_{z1}\hat{\sigma}_{z2}]\right\}, \tag{S11}$$

leading to
$$\begin{aligned}
\hat{H}_{\text{pol}}^I &= \int_0^\infty \frac{d\omega}{\sqrt{2\pi}}\sqrt{\kappa(\omega)}\left[(\hat{a} + \hat{\mathcal{O}})e^{-i\omega_r t} + (\hat{a}^\dagger + \hat{\mathcal{O}})e^{i\omega_r t}\right](\hat{b}_\omega e^{-i\omega t} + \hat{b}_\omega^\dagger e^{i\omega t}) \\
&\quad - \int_0^\infty \frac{d\omega}{\sqrt{2\pi}}\sqrt{\kappa(\omega)}(\hat{\mathcal{O}}e^{-i\omega_m t} + \hat{\mathcal{O}}e^{i\omega_m t})(\hat{b}_\omega e^{-i\omega t} + \hat{b}_\omega^\dagger e^{i\omega t}).
\end{aligned} \tag{S12}$$

This expression can be simplified by defining $\hat{B}(t) \equiv \int_0^\infty d\omega\,(\hat{b}_\omega e^{-i\omega t} + \hat{b}_\omega^\dagger e^{i\omega t})$ and assuming that $\kappa(\omega)$ is independent of frequency close to $\omega_m$ and $\omega_r$ according to the usual Markov approximation [S1]. Defining $\kappa(\omega_m) \equiv \kappa_m$ and $\kappa(\omega_r) \equiv \kappa_r$, we can write
$$\hat{H}_{\text{pol}}^I = \sqrt{\frac{\kappa_r}{2\pi}}\left[(\hat{a} + \hat{\mathcal{O}})e^{-i\omega_r t} + (\hat{a}^\dagger + \hat{\mathcal{O}})e^{i\omega_r t}\right]\hat{B}(t) - \sqrt{\frac{\kappa_m}{2\pi}}(\hat{\mathcal{O}}e^{-i\omega_m t} + \hat{\mathcal{O}}e^{i\omega_m t})\hat{B}(t) \tag{S13}$$
$$= \sum_n \sqrt{\frac{\kappa_n}{2\pi}}(\hat{\mathcal{C}}_n e^{-i\omega_n t} + \hat{\mathcal{C}}_n^\dagger e^{i\omega_n t})\hat{B}(t), \tag{S14}$$

where $n = r, m$ and $\hat{\mathcal{C}}_r \equiv \hat{a} + \hat{\mathcal{O}}, \hat{\mathcal{C}}_m \equiv -\hat{\mathcal{O}}$.

Using this result and following the standard approach [S1], we find a Born-Markov master equation

$$\dot{\rho}(t) = -\int_0^\infty d\tau \, \text{Tr}_f\{[\hat{H}^I_{\text{pol}}(t), [\hat{H}^I_{\text{pol}}(t-\tau), \rho(t) \otimes \rho_f]]\}, \tag{S15}$$

where $\rho(t)$ is the density matrix of the qubit-oscillator system and $\rho_f$ the density matrix of the oscillator's bath. Using Eq. (S13) and defining $S(t,\omega) \equiv \int_0^\infty d\tau \, \langle \hat{B}(t-\tau)\hat{B}(t)\rangle e^{-i\omega\tau}$, $S^*(t,\omega) \equiv \int_0^\infty d\tau \, \langle \hat{B}(t)\hat{B}(t-\tau)\rangle e^{i\omega\tau}$ results in

$$\begin{aligned}\dot{\rho}(t) = \sum_{n,n'} &- i[\Delta(\omega_n, -\omega_{n'}, t)\hat{\mathcal{C}}_n\hat{\mathcal{C}}_{n'}, \rho(t)] + \Gamma(\omega_n, -\omega_{n'}, t)\mathcal{D}[\hat{\mathcal{C}}_n, \hat{\mathcal{C}}_{n'}]\rho(t) \\ &- i[\Delta(-\omega_n, \omega_{n'}, t)\hat{\mathcal{C}}^\dagger_n\hat{\mathcal{C}}^\dagger_{n'}, \rho(t)] + \Gamma(-\omega_n, \omega_{n'}, t)\mathcal{D}[\hat{\mathcal{C}}^\dagger_n, \hat{\mathcal{C}}^\dagger_{n'}]\rho(t) \\ &- i[\Delta(-\omega_n, -\omega_{n'}, t)\hat{\mathcal{C}}^\dagger_n\hat{\mathcal{C}}_{n'}, \rho(t)] + \Gamma(-\omega_n, -\omega_{n'}, t)\mathcal{D}[\hat{\mathcal{C}}^\dagger_n, \hat{\mathcal{C}}_{n'}]\rho(t) \\ &- i[\Delta(\omega_n, \omega_{n'}, t)\hat{\mathcal{C}}_n\hat{\mathcal{C}}^\dagger_{n'}, \rho(t)] + \Gamma(\omega_n, \omega_{n'}, t)\mathcal{D}[\hat{\mathcal{C}}_n, \hat{\mathcal{C}}^\dagger_{n'}]\rho(t),\end{aligned} \tag{S16}$$

where

$$\mathcal{D}[\hat{x}, \hat{y}]\bullet = \hat{x} \bullet \hat{y} - \frac{1}{2}\{\hat{y}\hat{x}, \bullet\}, \tag{S17}$$

$$\Gamma(\omega_n, \omega_{n'}, t) = \frac{\sqrt{\kappa_n \kappa_{n'}}}{2\pi}[S(t, \omega_{n'}) + S^*(t, \omega_n)]e^{-i(\omega_n - \omega_{n'})t}, \tag{S18}$$

$$\Delta(\omega_n, \omega_{n'}, t) = \frac{i}{2}\frac{\sqrt{\kappa_n \kappa_{n'}}}{2\pi}[S(t, \omega_{n'}) - S^*(t, \omega_n)]e^{-i(\omega_n - \omega_{n'})t}. \tag{S19}$$

The last two equations correspond to the dissipation rates $\Gamma$ and the lamb shifts $\Delta$. To get an explicit form for $S(t,\omega)$, we consider a squeezed bath at a pump frequency $\omega_p$ [S1],

$$\langle \hat{b}_{\omega_1}\hat{b}_{\omega_2}\rangle = M(\omega_1)\delta(\omega_1 + \omega_2 - 2\omega_p), \tag{S20}$$

$$\langle \hat{b}^\dagger_{\omega_1}\hat{b}^\dagger_{\omega_2}\rangle = M^*(\omega_1)\delta(\omega_1 + \omega_2 - 2\omega_p), \tag{S21}$$

$$\langle \hat{b}^\dagger_{\omega_1}\hat{b}_{\omega_2}\rangle = N(\omega_1)\delta(\omega_1 - \omega_2), \tag{S22}$$

$$\langle \hat{b}_{\omega_1}\hat{b}^\dagger_{\omega_2}\rangle = [N(\omega_1) + 1]\delta(\omega_1 - \omega_2). \tag{S23}$$

Using these expressions in $S(t,\omega)$, we get two different expressions for $\omega > 0$ and $\omega < 0$

$$S(t, \omega > 0) = \pi\left[M(\omega)e^{-i2\omega_p t} + N(\omega) + 1\right], \tag{S24}$$

$$S(t, \omega < 0) = \pi\left[M^*(\omega)e^{i2\omega_p t} + N(\omega)\right]. \tag{S25}$$

Neglecting fast rotating terms in the usual secular approximation [S2], the dissipation rates take the form

$$\Gamma(\omega_n, -\omega_{n'}, t) = \frac{\sqrt{\kappa_n \kappa_{n'}}}{2}[M^*(\omega_n) + M^*(\omega_{n'})]e^{-i(\omega_{n'} + \omega_n - 2\omega_p)t}, \tag{S26}$$

$$\Gamma(-\omega_n, \omega_{n'}, t) = \frac{\sqrt{\kappa_n \kappa_{n'}}}{2}[M(\omega_n) + M(\omega_{n'})]e^{i(\omega_{n'} + \omega_n - 2\omega_p)t}, \tag{S27}$$

$$\Gamma(-\omega_n, -\omega_{n'}, t) = \frac{\sqrt{\kappa_n \kappa_{n'}}}{2}[N(\omega_{n'}) + N(\omega_n)]e^{i(\omega_n - \omega_{n'})t}, \tag{S28}$$

$$\Gamma(\omega_n, \omega_{n'}, t) = \frac{\sqrt{\kappa_n \kappa_{n'}}}{2}[N(\omega_{n'}) + N(\omega_n) + 2]e^{-i(\omega_n - \omega_{n'})t}, \tag{S29}$$

and the lamb shifts

$$\Delta(\omega_n, -\omega_{n'}, t) = \frac{i\sqrt{\kappa_n \kappa_{n'}}}{4}[M^*(\omega_{n'}) - M^*(\omega_n)]e^{-i(\omega_n + \omega_{n'} - 2\omega_p)t}, \tag{S30}$$

$$\Delta(-\omega_n, \omega_{n'}, t) = \frac{i\sqrt{\kappa_n \kappa_{n'}}}{4}[M(\omega_{n'}) - M(\omega_n)]e^{i(\omega_n + \omega_{n'} - 2\omega_p)t}, \tag{S31}$$

$$\Delta(-\omega_n, -\omega_{n'}, t) = \frac{i\sqrt{\kappa_n \kappa_{n'}}}{4}[N(\omega_{n'}) - N(\omega_n)]e^{i(\omega_n - \omega_{n'})t}, \tag{S32}$$

$$\Delta(\omega_n, \omega_{n'}, t) = \frac{i\sqrt{\kappa_n \kappa_{n'}}}{4}[N(\omega_{n'}) - N(\omega_n)]e^{-i(\omega_n - \omega_{n'})t}. \tag{S33}$$

See Sect. 6 for a discussion on the validity of the secular approximation here. In the next sections we consider three relevant cases: A) No squeezing: $N(\omega) = M(\omega) = 0$, B) squeezing at $\omega_p = (\omega_m + \omega_r)/2$ and C) squeezing at $\omega_p = \omega_r$ with a filter at $\omega_m$: $\kappa_m \to 0$.

## 1.1 No squeezing

The first case is a bath at zero temperature, corresponding to $M(\omega) = N(\omega) = 0$. The generalization to a finite temperature bath is simply obtained by setting $N(\omega) \neq 0$. The master equation Eq. (S16) then reduces to the Lindblad form

$$\dot{\rho}(t) = \mathcal{D}[\sqrt{\kappa_r}(\hat{a} + \hat{\mathcal{O}}) - \sqrt{\kappa_m}e^{i\delta t}\hat{\mathcal{O}}]\rho(t). \tag{S34}$$

where $\mathcal{D}[\hat{x}]\bullet = \hat{x} \bullet \hat{x}^\dagger - \frac{1}{2}\{\hat{x}^\dagger \hat{x}, \bullet\} = \mathcal{D}[\hat{x}, \hat{x}^\dagger]\bullet$ is the usual dissipation superoperator. In this frame, the oscillator starts and stays in vacuum which means that the terms $\mathcal{D}[a, \hat{\mathcal{O}}]\rho(t)$ and $\mathcal{D}[\hat{\mathcal{O}}, \hat{a}^\dagger]\rho(t)$ will be zero at all times. We can thus rewrite the equation in a way that makes the qubit dephasing rate explicit

$$\dot{\rho}(t) = \kappa_r \mathcal{D}[\hat{a}]\rho(t) + [\kappa_r + \kappa_m - 2\sqrt{\kappa_r\kappa_m}\cos(\delta t)]\mathcal{D}[\hat{\mathcal{O}}]\rho(t). \tag{S35}$$

Moving out of the interaction picture and setting $g_1 = g_2 \equiv g$, $\kappa_r = \kappa_m \equiv \kappa$, $1/\tilde{\delta} \approx 1/\delta$ for simplicity, we recover Eq. (4) of the main paper. We also see from this equation that we can use a filter at the modulation frequency to lower the dephasing rate. Setting $\kappa_m \to 0$, we get $\Gamma = \kappa(g/2\delta)^2$ which is on average a factor two decrease over the initial dephasing rate.

## 1.2 Squeezing at the average frequency

The second case we consider is a bath with a broadband squeezing spectrum centred at the average of the oscillator and modulation frequency $\omega_p = (\omega_r + \omega_m)/2$. We assume a flat squeezing spectrum over the relevant bandwidth such that $M \equiv M(\omega_m) = M(\omega_r)$ and $N \equiv N(\omega_m) = N(\omega_r)$. In the limit of perfect squeezing, we can write $M = \sqrt{N(N+1)}e^{2i\theta}$ with $\theta$ the squeezing angle and $N = \sinh^2 r$ with $r$ the squeezing parameter. As explained in the main paper, the condition on $\theta$ is that the anti-squeezed quadrature is aligned with the displacement direction at all time. Note that here we set the displacement direction by fixing the phase reference of the first qubit modulation drive, so that $\theta$ is also referenced to the modulation. Setting $\theta = 0$ and assuming a flat spectrum in the output field density of modes $\kappa_r = \kappa_m \equiv \kappa$, we get after some algebra that Eq. (S16) can be written in Lindblad form

$$\dot{\rho}(t) = \kappa\, \mathcal{D}\left[\cosh(r)\hat{a}e^{-\frac{i\delta t}{2}} + \sinh(r)\hat{a}^\dagger e^{\frac{i\delta t}{2}} - ie^{-r}\sin\left(\frac{\delta t}{2}\right)\hat{\mathcal{O}}\right]\rho(t). \tag{S36}$$

The last term in the dissipation operator clearly shows that phase information is completely hidden at high squeezing power. Assuming that the squeezing interaction has been turned on long before the gate, the oscillator starts in a squeezed vacuum state ($\alpha_j(0) = 0$ in Eq. (S5)). In the polaron frame, the oscillator is at all times in a squeezed vacuum state, which means that the dephasing rate is given by the prefactor in front of the qubit ($\hat{\mathcal{O}}$) operator

$$\Gamma = 2\kappa\left(\frac{g}{2\tilde{\delta}}\right)^2[1 - \cos(\delta t)]e^{-2r}, \tag{S37}$$

for $g_1 = g_2 \equiv g$.

This equation indicates that we dephasing can be reduced exponentially for arbitrarily high squeezing levels $r$, but one must keep in mind the approximations that were made in order to get the final master equation Eq. (S36). In particular, we neglected fast-rotating terms in the secular approximation when calculating the dephasing rates of the master equation, Eq. (S26). Since $N$ and $M$ grow exponentially with squeezing power, there will be a point where the secular approximation is no longer valid. In general the error made due to the secular approximation can be upper bounded by

$$\varepsilon_{\text{secular}} \lesssim \frac{2}{3}\left[2\kappa\left(\frac{g}{2\tilde{\delta}}\right)^2 e^{2r}\right]^2 t_g \frac{2\pi}{(2\omega_m)}, \tag{S38}$$

following the approach outlined in Sect. 6. The physical intuition behind this error bound is that the real path in phase space of the oscillator, given by $\alpha(t)$ in Eq. (S6), is not a perfect circle due to the fast rotating terms. Thus, the anti-squeezed quadrature cannot be aligned with the displacement at all times and small deviations from the circle will eventually lead to an increase in dephasing at very high squeezing power.

For the parameters used in Fig. 2 (c) of the main paper, $\delta/(2\pi) = 0.6$ GHz, $g/(2\pi) = 60$ MHz, $t_g = 42.7$ ns, $\kappa/(2\pi) = 1.0$ MHz and $\omega_r/(2\pi) = 10$ GHz, the right-hand side of Eq. (S38) evaluates

to $\sim 10^{-5}$ for $S = 10$ dB of squeezing. Hence we expect that we cannot be confident about a gate error smaller than this number based on evaluating Eq. (S36) with this set of parameters. The curve in Fig. 3 of the paper is only shown for values of $S$ smaller than this bound.

We also assumed an equal squeezing spectrum and equal decay rates at the two frequencies $\omega_r$ and $\omega_m$. It is not shown in Eq. (S36), but a discrepancy in the decay rates and/or squeezing spectrum will lead to an additional dephasing that grows exponentially with squeezing.

### 1.3 Squeezing at the oscillator frequency

The third case we consider is a bath squeezed at the resonator frequency, $\omega_p = \omega_r$. We set $N(\omega_m) = M(\omega_m) = 0$ and define $M \equiv M(\omega_r)$, $N \equiv N(\omega_r)$, which corresponds to a squeezing spectrum much larger than $\kappa_r$, but much narrower than $\delta$. Furthermore, we will assume a filter at the oscillator's output so that $\kappa_m = 0$, leading to

$$\dot{\rho}(t) = \kappa_r(N+1)\mathcal{D}[\hat{a} + \hat{\mathcal{O}}]\rho(t) + \kappa_r N \mathcal{D}[\hat{a}^\dagger + \hat{\mathcal{O}}]\rho(t) + \kappa_r M \mathcal{S}[\hat{a}^\dagger + \hat{\mathcal{O}}]\rho(t) + \kappa_r M^* \mathcal{S}[\hat{a} + \hat{\mathcal{O}}]\rho(t). \tag{S39}$$

In this expression, $\mathcal{S}[\hat{x}]\bullet = \hat{x} \bullet \hat{x} - \frac{1}{2}\{\hat{x}\hat{x}, \bullet\} = \mathcal{D}[\hat{x}, \hat{x}]\bullet$ is a squeezing superoperator. The gain in fidelity appears when we rewrite this in Lindblad form and set $\theta = \pi/2$

$$\dot{\rho}(t) = \kappa_r \mathcal{D}[\cosh(r)\hat{a} - \sinh(r)\hat{a}^\dagger + e^{-r}\hat{\mathcal{O}}]\rho(t). \tag{S40}$$

The dephasing rate is then exponentially reduced

$$\Gamma = \kappa_r \left(\frac{g}{2\tilde{\delta}}\right)^2 e^{-2r}, \tag{S41}$$

for $g_1 = g_2 \equiv g$. Phase information is completely hidden for large squeezing.

Since this equation is derived the same way as Eq. (S36), this exponential gain also has a confidence bound similar to Eq. (S38). We also note that adding filter this way does no contradict the Markov approximation made earlier. As long as the filter's bandwith is smaller than $\delta$, making the Markov approximation here amounts to assuming that the bath density of modes is constant around each frequency in play ($\omega_r$ and $\omega_m$).

## 2 Examples of physical implementation

In this section we give a very brief summary of how we can achieve a longitudinal coupling modulation in various platforms, as well as references on how to perform single qubit control.

*Transmons*—More details about transmons longitudinally-coupled to resonators can be found in [S3–S5]. Arbitrary single qubit $X = \hat{\sigma}_x$ and $Y = \hat{\sigma}_y$ rotations can be performed in the standard way by applying a microwave drive at a side gate voltage. Fidelities for these gates are now above 99.9% [S6]. The longitudinal coupling $g_z(t)$ can be modulated via an AC flux drive $\Phi_x(t)$ in the middle of the qubit squid loop: The frequency and the amplitude of the coupling modulation are then directly related to the frequency and amplitude of the flux drive.

*Flux qubits*—In this implementation, for which more details can be found in Ref. [S7], X and Y single-qubit gates can be realized via modulation of the flux inside the qubit loop with over 99.8% fidelity [S8]. Modulation of the longitudinal coupling can be realized via modulation of the reduced gate charge on the superconducting island.

*Spin qubits*—As discussed in more details in [S9], in this implementation the longitudinal coupling could be modulated by controlling the inter-dot tunnelling. Single-qubit gates with average fidelities of 99.6% have been demonstrated [S10].

*Singlet-triplet spin qubits*—For this implementation, single qubit operations with 99% have been demonstrated [S11]. Similar to the previous implementation, longitudinal coupling to a resonator can be modulated through the inter-dot tunnelling [S12].

## 3 Coupled oscillators

In this section we derive the effective $\hat{\sigma}_z\hat{\sigma}_z$-coupling induced when the two qubits are in different but coupled oscillators. Similar to the single oscillator case, both couplings are modulated at the same

frequency $\omega_m$. The Hamiltonian corresponding to this situation with capacitively coupled oscillators is

$$\hat{H} = \omega_a \hat{a}^\dagger \hat{a} + \omega_b \hat{b}^\dagger \hat{b} + \tfrac{1}{2}\omega_{a1}\hat{\sigma}_{z1} + \tfrac{1}{2}\omega_{a2}\hat{\sigma}_{z2} + g_1(t)\hat{\sigma}_{z1}(\hat{a}^\dagger + \hat{a}) + g_2(t)\hat{\sigma}_{z2}(\hat{b}^\dagger + \hat{b}) - g_{ab}(\hat{a}^\dagger - \hat{a})(\hat{b}^\dagger - \hat{b}). \tag{S42}$$

Following the same approach as in the single oscillator case, we first move to a frame rotating at $\omega_m$ for both oscillators and $\omega_{ai}$ for the respective qubits. To simplify the discussion, we also perform a rotating wave approximation and neglect fast-rotating terms leading to

$$\hat{H}_R = \delta_a \hat{a}^\dagger \hat{a} + \delta_b \hat{b}^\dagger \hat{b} + \frac{g_1}{2}\hat{\sigma}_{z1}(\hat{a}^\dagger + \hat{a}) + \frac{g_2}{2}\hat{\sigma}_{z2}(\hat{b}^\dagger + \hat{b}) + g_{ab}(\hat{a}^\dagger \hat{b} + \hat{a}\hat{b}^\dagger). \tag{S43}$$

The second step is to diagonalize the oscillator part of the Hamiltonian $\hat{H}_r = \delta_a \hat{a}^\dagger \hat{a} + \delta_b \hat{b}^\dagger \hat{b} + g_{ab}(\hat{a}^\dagger \hat{b} + \hat{a}\hat{b}^\dagger)$ and to express the longitudinal coupling in terms of the resulting hybridized modes. For this purpose, we define the eigenmode operators

$$\hat{c} = \cos\xi\,\hat{a} + \sin\xi\,\hat{b}, \tag{S44}$$
$$\hat{d} = -\sin\xi\,\hat{a} + \cos\xi\,\hat{b}, \tag{S45}$$

where $\tan 2\xi = 2g_{ab}/(\omega_a - \omega_b)$. Expressing the Hamiltonian Eq. (S43) in terms of these eigenmodes yields

$$\hat{H}_R = \delta_c \hat{c}^\dagger \hat{c} + \delta_d \hat{d}^\dagger \hat{d} + \left\{ \frac{1}{2}[g_1 \cos\xi\,\hat{\sigma}_{z1} + g_2 \sin\xi\,\hat{\sigma}_{z2}]\hat{c}^\dagger + \frac{1}{2}[-g_1 \sin\xi\,\hat{\sigma}_{z1} + g_2 \cos\xi\,\hat{\sigma}_{z2}]\hat{d}^\dagger + \text{H.c.} \right\}, \tag{S46}$$

where the detunings are

$$\delta_c = \frac{\delta_a + \delta_b}{2} + \frac{g_{ab}}{\sin 2\xi}, \qquad \delta_d = \frac{\delta_a + \delta_b}{2} - \frac{g_{ab}}{\sin 2\xi}. \tag{S47}$$

Following the same approach as in Sect. 1 we finally apply the unitary transformation

$$\hat{U}_D = e^{\hat{\mathcal{O}}_c \hat{c}^\dagger - \hat{\mathcal{O}}_c^\dagger \hat{c}} e^{\hat{\mathcal{O}}_d \hat{d}^\dagger - \hat{\mathcal{O}}_d^\dagger \hat{d}} \equiv \hat{U}_c \hat{U}_d, \tag{S48}$$

with $\hat{\mathcal{O}}_c = (g_1 \cos\xi\,\hat{\sigma}_{z1} + g_2 \sin\xi\,\hat{\sigma}_{z2})/2\delta_c$ and $\hat{\mathcal{O}}_d = (-g_1 \sin\xi\,\hat{\sigma}_{z1} + g_2 \cos\xi\,\hat{\sigma}_{z2})/2\delta_d$. Because $[\hat{U}_c, \hat{U}_d] = 0$, the transformation does not generate a coupling between the eigenmodes. The resulting Hamiltonian is then

$$\hat{H}_{\text{pol}} = \delta_c \hat{c}^\dagger \hat{c} + \delta_d \hat{d}^\dagger \hat{d} + J_z(t)\hat{\sigma}_{z1}\hat{\sigma}_{z2}, \tag{S49}$$

with the $\hat{\sigma}_z \hat{\sigma}_z$-coupling strength

$$\bar{J}_z = \frac{\delta_c - \delta_d}{\delta_c \delta_d} \frac{g_1 g_2}{4} \sin 2\xi. \tag{S50}$$

Defining $\bar{\delta} = (\delta_a + \delta_b)/2$ and $\zeta = 1/\tan 2\xi$, we can write this as

$$\bar{J}_z = \frac{1}{2} \frac{g_1 g_2 g_{ab}}{\bar{\delta}^2 - g_{ab}^2(1 + \zeta^2)} \tag{S51}$$

corresponding to the result stated in the main paper.

## 4 Details on the simulations

To calculate the average gate fidelity, we compare full master equation simulations to the ideal channel $\mathcal{U}_{CZ}\bullet = \hat{U}_{CZ} \bullet \hat{U}_{CZ}^\dagger$ with $\hat{U}_{CZ} = \text{diag}[1,1,1,-1] = \hat{U}_{CP}(\pi)$. The master equation is defined over the system oscillator-qubit and we therefore eliminate the oscillator degree of freedom. We define the channel $\mathcal{E}_{q_1 q_2}$ acting on the qubits as

$$\mathcal{E}_{q_1 q_2}(\bullet) = \mathcal{T}_r\,\mathcal{U}_D^\dagger\,\mathcal{U}_I^\dagger\,\mathcal{E}^{t_g}\,\mathcal{U}_I\,\mathcal{U}_D\,\mathcal{S}\left(|0\rangle\langle 0|_r \otimes \bullet\right), \tag{S52}$$

where the superoperator $\mathcal{T}_r\bullet \equiv \text{Tr}_r(\bullet)$ is the trace over the oscillator degree of freedom and the superoperators $\mathcal{U}_D\bullet \equiv \hat{\mathbf{U}}\bullet\hat{\mathbf{U}}^\dagger$, $\mathcal{U}_D^\dagger\bullet \equiv \hat{\mathbf{U}}^\dagger\bullet\hat{\mathbf{U}}$ are the unitary displacement transformations defined in Eq. (S5). We also defined the interaction picture superoperators $\mathcal{U}_I\bullet \equiv \hat{U}_I \bullet \hat{U}_I^\dagger$ and $\mathcal{U}_I^\dagger\bullet \equiv \hat{U}_I^\dagger \bullet \hat{U}_I$

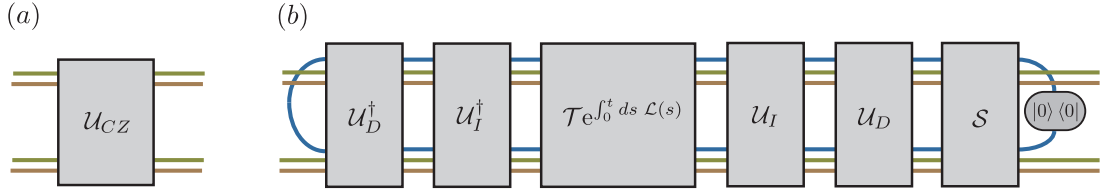

Figure S1: Tensor representation of the CZ gate supermatrix [S13]. Here the blue lines represent the oscillator degree of freedom and the green (brown) lines represent the first (second) qubit. (a) Ideal controlled-Z channel. (b) Simulated quantum channel over two qubits. We project onto an initial vacuum state of the oscillator at the beginning (right) and trace over the oscillator degree of freedom at the end (left). The superoperator $\mathcal{T}e^{\bullet}$ denotes the time-ordering exponential.

with $\hat{U}_I$ defined in Eq. (S11). The squeezing superoperator is given by $\mathcal{S} \equiv \hat{\mathbf{S}}(re^{2i\theta})^\dagger \bullet \hat{\mathbf{S}}(re^{2i\theta})$ with the standard definition $\hat{\mathbf{S}}(re^{2i\theta}) = \exp\left[re^{-2i\theta}\hat{a}^2/2 - \text{H.c.}\right]$.

Finally, $\mathcal{E}^{t_g}$ is the oscillator-qubit channel calculated from numerical integration of the differential equation

$$\dot{\mathcal{E}}^t = \mathcal{L}\mathcal{E}^t. \tag{S53}$$

where $\mathcal{L}$ is a Liouvillian derived in Sect. 1. If we impose the initial condition $\mathcal{E}^0 = \text{Id}$, then $\mathcal{E}^t$ denotes the channel resulting from evolution under the Liouvillian $\mathcal{L}$ for a time $t$. This whole procedure is illustrated in terms of tensor diagrams in Fig. S1 [S13].

In the case where we include intrinsic qubit decay and dephasing, we add additional terms to the calculated Liouvillian

$$\dot{\rho}(t) = \mathcal{L}'\rho(t) = \mathcal{L}\rho(t) + \sum_i \gamma_1^{(i)}\mathcal{D}[\hat{\sigma}_{-i}]\rho(t) + \gamma_\phi^{(i)}\mathcal{D}[\hat{\sigma}_{zi}]\rho(t), \tag{S54}$$

where the decay rates are given by $\gamma_1^{(i)} = 1/T_1^{(i)}$, $1/T_2^{(i)} = \gamma_\phi^{(i)} + \gamma_1^{(i)}/2$.

Although we take the oscillator to initially be in a vacuum squeezed state in our numerical calculations, we emphasize that any initial displacement could have been added without change to the resulting fidelity.

Knowing the effective channel over two qubits, the average gate fidelity $\mathcal{F}$ is obtained by averaging over all two-qubit initial states according to the uniform (Haar) measure [S14].

$$\mathcal{F} = \int d\psi \langle \psi | U_{CZ}^\dagger \mathcal{E}_{q_1 q_2}(|\psi\rangle\langle\psi|) U_{CZ} |\psi\rangle. \tag{S55}$$

The simulations were performed using QuTiP [S15].

## 5 Comparison with transverse resonator-induced phase gate

In the main Paper, we present a numerical example for the gate time and gate fidelity using $\kappa/2\pi = 0.05$ MHz [S16], $g/2\pi = 60$ MHz [S3] and $\delta/2\pi = 537$ MHz. This results in a very short gate time of $t_g = 37$ ns with an average gate infidelity as small as $1 \times 10^{-4}$. Taking into account finite qubit lifetimes $T_1 = 30\,\mu\text{s}$ and $T_2 = 20\,\mu\text{s}$ [S6], we find that the infidelity is increased to $\sim 10^{-3}$. As also pointed out in the Paper, the gate fidelity is limited by the qubit's natural decoherence channels with these parameters.

For the same value of $\kappa$ and typical circuit QED parameters, ideal simulations (excluding $T_1$ and $T_2$) of a transversal resonator-induced phase (RIP) gates yield a gate fidelity of $4 \times 10^{-4}$ for a gate time of 200 ns [S17]. Thus, comparing to a transversal RIP gate with parameters from Ref. [S17], the scheme introduced here, with the representative choice of parameters used in the previous paragraph, exhibits a factor 4 improvement in fidelity and a factor 5 improvement in gate time. Since a transversal RIP gate depends on a different set of parameters (in particular there is a strong dependence on resonator-qubit detuning), the use of optimal control would allow the comparison of best-case performance for the gate proposed in this paper and transversal RIP gates. The large improvement we find for typical parameter choices and un-optimized pulse shapes, however, suggests that very substantial improvements are possible in practice.

# 6 Error bound for rotating terms in the master equation

In this section, we estimate an upper bound on the error made by neglecting fast-rotating terms in the master equation. More precisely, we want to estimate where the secular approximation made in Eqs. (S26) to (S33) is no longer valid. We will follow closely the supplemental material of [S18], where a similar question was addressed for unitary evolution.

We start with a general Lindbladian $\mathcal{L}(t) = \mathcal{L}_0 + \gamma(t)\mathcal{L}_1$ and we will assume that $\mathcal{L}_0, \mathcal{L}_1$ do not depend on time and that $\gamma(t)$ is some fast-oscillating function. In particular, we are interested in the case where

$$\int_0^{\Delta t} dt\, \gamma(t) = 0, \tag{S56}$$

with $\Delta t$ the smallest time increment for which Eq. (S56) is respected. In our case, we have $\Delta t \sim 2\pi/(2\omega_m)$.

The problem we address is the following: What is the error we make when we replace $\mathcal{L}(t)$ by

$$\mathcal{L}_{av} \equiv \frac{1}{\Delta t}\int_0^{\Delta t} dt\, \mathcal{L}(t) = \mathcal{L}_0. \tag{S57}$$

In other words, what is the error we make by doing a Suzuki-Trotter decomposition and at each time step we replace the linbladian by its average. We define an average channel $\mathcal{E}_{av}(t) \equiv e^{\mathcal{L}_{av}t}$, which replaces the full evolution channel $\mathcal{E}(t) = \mathcal{T}e^{\int_0^t ds\, \mathcal{L}(s)}$. To estimate the error for a single time step, we look at the norm of the superoperator $X(\Delta t) \equiv \mathbb{I} - \mathcal{E}_{av}^{-1}(\Delta t)\mathcal{E}(\Delta t)$. Knowing that $X(0) = 0$, we write

$$\begin{aligned}
X(\Delta t) &= \int_0^{\Delta t} ds\, \dot{X}(s) \\
&= -\int_0^{\Delta t} ds\, \dot{\mathcal{E}}_{av}^{-1}(s)\mathcal{E}(s) + \mathcal{E}_{av}^{-1}(s)\dot{\mathcal{E}}(s) \\
&= \int_0^{\Delta t} ds\, \mathcal{E}_{av}^{-1}(s)\mathcal{L}_{av}\mathcal{E}(s) - \mathcal{E}_{av}^{-1}(s)\mathcal{L}(s)\mathcal{E}(s)
\end{aligned} \tag{S58}$$

where we used the differential equation for the channel $\dot{\mathcal{E}} = \mathcal{L}\mathcal{E}$ and we directly differentiated $\dot{\mathcal{E}}_{av}^{-1} = \partial_t(e^{-\mathcal{L}_{av}t}) = -\mathcal{E}_{av}^{-1}\mathcal{L}_{av}$. We replace $\mathcal{L}_{av}$ by its explicit expression Eq. (S57) and change the integration variables to get

$$X(\Delta t) = \frac{1}{\Delta t}\int_0^{\Delta t} ds \int_0^{\Delta t} d\tau\, \mathcal{E}_{av}^{-1}(s)\mathcal{L}(\tau)\mathcal{E}(s) - \mathcal{E}_{av}^{-1}(\tau)\mathcal{L}(\tau)\mathcal{E}(\tau). \tag{S59}$$

We now evaluate the norm and use the triangle inequality to get

$$\|X(\Delta t)\| \leqslant \frac{1}{\Delta t}\int_0^{\Delta t} ds \int_0^{\Delta t} d\tau\, \left\|\mathcal{E}_{av}^{-1}(\tau)\mathcal{L}(\tau)\left[\mathcal{E}(s) - \mathcal{E}(\tau)\right]\right\| + \left\|\left[\mathcal{E}_{av}^{-1}(s) - \mathcal{E}_{av}^{-1}(\tau)\right]\mathcal{L}(\tau)\mathcal{E}(\tau)\right\|. \tag{S60}$$

We know that a physical channel is norm contractive and we will choose a norm respecting $\|\mathcal{E}\| \leq 1$ so that we can use the Schwartz inequality and write

$$\|X(\Delta t)\| \leqslant \frac{\|\mathcal{L}\|}{\Delta t}\int_0^{\Delta t} ds \int_0^{\Delta t} d\tau\, \left\|\mathcal{E}_{av}^{-1}(\tau)\right\|\|\mathcal{E}(s) - \mathcal{E}(\tau)\| + \left\|\mathcal{E}_{av}^{-1}(s) - \mathcal{E}_{av}^{-1}(\tau)\right\|, \tag{S61}$$

where we defined $\|\mathcal{L}\| = \max_s \|\mathcal{L}(s)\|$. The first term in the integral is upper bounded by

$$\begin{aligned}
\|\mathcal{E}(s) - \mathcal{E}(\tau)\| &= \left\|\int_s^\tau dt\, \dot{\mathcal{E}}(t)\right\| \\
&= \left\|\int_\tau^s dt\, \mathcal{L}(t)\mathcal{E}(t)\right\| \\
&\leqslant \|\mathcal{L}\|\,|s - \tau|,
\end{aligned} \tag{S62}$$

and the second term by

$$\begin{aligned}\left\|\mathcal{E}_{av}^{-1}(s)-\mathcal{E}_{av}^{-1}(\tau)\right\| &= \left\|\int_s^\tau dt\, \dot{\mathcal{E}}_{av}^{-1}(t)\right\| \\ &= \left\|\mathcal{L}_{av}\int_s^\tau dt\, \mathcal{E}_{av}^{-1}(t)\right\| \\ &\leqslant \|\mathcal{L}\|\left\|\mathcal{E}_{av}^{-1}(\Delta t)\right\| |s-\tau|. \end{aligned} \quad (S63)$$

In the last line we used that fact that if a physical channel contracts the norm, then its inverse must necessarily increase it which means that for $t_1 > t_2$ we have $\left\|\mathcal{E}^{-1}(t_1)\right\| \geq \left\|\mathcal{E}^{-1}(t_2)\right\|$ Putting back Eqs. (S62) and (S63) into Eq. (S61), we get

$$\begin{aligned}\|X(\Delta t)\| &\leqslant \frac{2\|\mathcal{L}\|^2}{\Delta t}\left\|\mathcal{E}_{av}^{-1}(\Delta t)\right\|\int_0^{\Delta t} ds \int_0^{\Delta t} d\tau\, |s-\tau| \\ &\leqslant \frac{2}{3}\|\mathcal{L}\|^2 \Delta t^2 \left\|\mathcal{E}_{av}^{-1}(\Delta t)\right\| \\ &\lesssim \frac{2}{3}\|\mathcal{L}\|^2 \Delta t^2. \end{aligned} \quad (S64)$$

In general, the norm of the average inverse channel can be large, but in our specific case $\Delta t$ is much smaller than any evolution time scale of the average channel, which means that we can approximate $\left\|\mathcal{E}_{av}^{-1}(\Delta t)\right\| \approx 1$. Knowing the error made for each $\Delta t$ step, we get an upper bound for the full evolution

$$\|X(t_g)\| \lesssim \frac{2}{3}\|\mathcal{L}\|^2 t_g \Delta t. \quad (S65)$$

We add that in order to apply this bound in a meaningful way, the operators inside $\mathcal{L}$ should be bounded, which is not the case for the oscillator operators $\hat{a}, \hat{a}^\dagger$. However, using a Bogoliubov transformation followed by the steps used to go from Eq. (S34) to Eq. (S35) we can express the master equations Eqs. (S36) and (S40) in a form similar to Eq. (S35) where we can trace out the harmonic oscillator. That way, the bound Eq. (S65) can be applied on an effective two-qubit master equation where all the operators are bounded.



\end{document}